\documentclass[12pt]{article}
\usepackage{amstex}

\columnwidth\textwidth

\begin{document}

\newcommand{\be}{\begin{equation}}
\newcommand{\ee}{\end{equation}}
\newcommand{\beann}{\begin{eqnarray*}}
\newcommand{\eeann}{\end{eqnarray*}}
\newcommand{\bea}{\begin{eqnarray}}
\newcommand{\eea}{\end{eqnarray}}
\newcommand{\nn}{\nonumber}
\newcommand{\ben}{\begin{enumerate}}
\newcommand{\een}{\end{enumerate}}
\newtheorem{df}{Definition}
\newtheorem{thm}{Theorem}
\newtheorem{lem}{Lemma}
\newtheorem{prop}{Proposition}
\begin{titlepage}

\noindent
\hspace*{11cm} BUTP-01/26 \\
\vspace*{1cm}
\begin{center}
{\LARGE Pair of null gravitating shells III. \newline
Algebra of Dirac's observables} 

\vspace{2cm}

I.~Kouletsis and P. H\'{a}j\'{\i}\v{c}ek \\
Institute for Theoretical Physics \\
University of Bern \\
Sidlerstrasse 5, CH-3012 Bern, Switzerland \\
\vspace*{2cm}

December 2001 \\ \vspace*{1cm}

\nopagebreak[4]

\begin{abstract}
  The study of the two-shell system started in ``Pair of null gravitating
  shells I and II'' is continued. The pull back of the Liouville form to the
  constraint surface, which contains complete information about the Poisson
  brackets of Dirac observables, is computed in the singular double-null
  Eddington-Finkelstein (DNEF) gauge. The resulting formula shows that the
  variables conjugate to the Schwarzschild masses of the intershell spacetimes
  are simple combinations of the values of the DNEF coordinates on these
  spacetimes at the shells.  The formula is valid for any number of in- and
  out-going shells. After applying it to the two-shell system, the symplectic
  form is calculated for each component of the physical phase space; regular
  coordinates are found, defining it as a symplectic manifold. The symplectic
  transformation between the initial and final values of observables for the
  shell-crossing case is written down.

\end{abstract}

\end{center}

\end{titlepage}

\section{Introduction}
\label{sec:intro}
The present paper is the third in a series devoted to the two-shell system.
The first paper, Ref.~\cite{I}, and the second, Ref.~\cite{II}, will be 
referred to as I and II. In I, all classical solutions that have a regular
center on the left are described, and the
space of solutions is parametrized by three discrete and four continuous
parameters.  The space of solutions is a candidate for the physical phase
space and the parameters are candidates for Dirac observables. In II, the
action functional for a single shell due to Louko, Whiting and Friedman 
\cite{L-W-F} is
generalized to any number of in- and out-going shells. The pull back
$\Theta_\Gamma$ of the corresponding Liouville form $\Theta$ to the constraint
surface $\Gamma$ is transformed into coordinates consisting of embeddings,
embedding momenta, and Dirac observables. Some general properties of the pull
back, such as gauge invariance, are shown. They enable us to accomplish
the transformation explicitly, and this is the main task of the present paper.

The central idea of this paper is to calculate $\Theta_\Gamma$ in double-null
Eddington-Finkelstein (DNEF) coordinates. This is a kind of gauge, but a {\em
  singular} one: Both the metric and the embeddings corresponding to this
gauge are discontinuous at the shells and diverging at any Schwarzschild
horizon. We shall show that this singularity does not influence the calculated
value of $\Theta_\Gamma$. That follows from the way $\Theta_\Gamma$ transforms
under ordinary gauge transformations. What is the motivation for using the
singular gauge?

In fact, we first calculated it in regular gauges. The results
revealed that the DNEF coordinates had a special role: their values at
the shells appeared in the final formula for
$\Theta_\Gamma$. Moreover, the calculations were very long while the
final formula was very simple, indicating that working with DNEF
coordinates from the start could lead to simplifications. This turns
out to be right, although the calculation is still far from being
trivial.  One can perhaps say that the Liouville form chooses itself
the spacetime coordinates in which it likes to be expressed.

Originally, $\Theta_\Gamma$ has a form of an integral over a Cauchy
surface.  Each Cauchy surface has a boundary consisting of the regular
center and the infinity; each such surface intersects the shells at
{\em intersection points}.  Due to the discontinuity of the DNEF gauge
at the shells, the contribution of each intersection point to
$\Theta_\Gamma$ is non-zero, whereas the contributions from the
regular center and the infinity in this gauge both vanish
(cf.~\cite{H-Kie}, where the only non-zero contribution comes from the
infinity). Thus, $\Theta_\Gamma$ can be cast as a sum over all
intersection points in which the summands have a standard form. In
this way, a general formula can be shown to be valid for any number of
in- or out-going shells.

The sum over the intersection points is of course associated with a
particular Cauchy surface. Let us consider two such surfaces.  If the
shells do not cross between these two Cauchy surfaces, then the
corresponding summands have the same value on each surface. However,
if the shells do cross between the surfaces, the corresponding
summands for the Cauchy surface below and above the crossing point are
related by a highly non-trivial canonical transformation.

The plan of the paper is as follows. In Sec.~\ref{sec:liouville}, the
calculation of $\Theta_\Gamma$ in the DNEF gauge is justified and
accomplished. The result is a general formula that expresses $\Theta_\Gamma$
in terms of some Dirac observables. These are the Schwarzschild masses of the
intershell spacetime pieces and some combinations of the values of DNEF
coordinates on these pieces at the shells. The canonical
transformation between the observables below and above the crossing point 
is calculated.

In Sec.~\ref{sec:algebra}, the formula is specialized to the system of
two shells. Some complete sets of Dirac observables as coordinates on
the various components and regions of the physical phase space are
considered.  The physical phase space is then given the structure of a
symplectic manifold. On each component of the physical phase space, we
find a global chart with respect to which the components of the
symplectic form are $C^\infty$ and regular, in the sense that the
matrix of the components has a nowhere vanishing determinant. In
particular, in the case of shell-crossing, the ``singular'' case in
which both shells lie on horizons (denoted by C$_{00}$ in I) turns out
to be a smooth surface in the phase space. Finally,
Sec.~\ref{sec:outlook} contains some conclusions and outlook,
speculating about the prospective quantum theory.

\section{Calculation of the Liouville form in a singular gauge}
\label{sec:liouville} Let $\Sigma$ be a Cauchy surface defined by an
embedding $(U(\rho),V(\rho))$.  In II, the pull back $\Theta_\Gamma$
to the constraint surface $\Gamma$ of the Liouville form $\Theta$ has
been written as a sum of various contributions from different parts of
$\Sigma$: First, there are contributions from each spacetime point
(denoted by $p$) where a shell intersects $\Sigma$. If the
intersection is with a single (in-going or out-going) shell, each
point $p$ contributes by a single term (cf.~II, Eq.~(46)) 
\[
  {\mathbf p}d{\mathbf r}\ .  
\] 
If the intersection is a point $p$
where an in-going and an out-going shells cross each other, then the
contribution is 
\[
  {\mathbf p}_{\text{out}}d{\mathbf r} + 
  {\mathbf p}_{\text{in}}d{\mathbf r}\ .
\]
Second, there are contributions from each connected volume cut out
from $\Sigma$ by the shell intersections. Each such volume contributes
by the boundary terms 
\[
  (fdU + gdV + h_ido^i - \varphi db)_{\rho=b} - (fdU + gdV + h_ido^i - \varphi
  da)_{\rho=a} \ ,
\]
where $\rho = a$ and $\rho = b$, $a < b$, are the boundary points of the
volume (cf.~Eq.~(60) of II), $o^i$ are Dirac observables, $f$, $g$, $h_i$ and
$\varphi$ are functions of $o^i$, $U(\rho)$ and $V(\rho)$ defined by
Eqs.~(51)--(55) of II:
\begin{eqnarray}
f &=& \frac{RR_{,U}}{2}\ln\left(-\frac{U'}{V'}\right)
 +F(U,V,o^i)\ ,
\label{6,1} \\
g &=& \frac{RR_{,V}}{2}\ln\left(-\frac{U'}{V'}\right)+G(U,V,o^i)\ ,
\label{6,2}\\
h_i &=& \frac{RR_{,i}}{2}\ln\left(-\frac{U'}{V'}\right)
 +H_i(U,V,o^i)\ ,
\label{6,3} \\
 \varphi &=& RR_{,U}U'-RR_{,V}V'-\frac{R}{2}(R_{,U}U'+R_{,V}V')
 \ln\left(-\frac{U'}{V'}\right)\\ & & \; -FU'-GV'+\phi(U,V,o^i)\ ,
\label{6,4} 
\end{eqnarray} 
The boundaries $a$ and $b$ can correspond either to the regular center
$\rho = 0$ of $\Sigma$, or to a shell intersection $\rho = {\mathbf
r}$ with $\Sigma$, or to the infinity $\rho = \infty$ of
$\Sigma$. Finally, there is a contribution due to the infinity,
\[
  -{\mathcal N}_\infty E_\infty dt
\]
(cf.~Eq.(46) of II).

In this paper, we shall collect all contributions at each particular
boundary and transform it in several steps to a general simple form. We shall
denote the contribution from the center by $\Theta_0$, from any shell
intersection point $\mathbf r$ by $\Theta_{\mathbf r}$ and from the infinity
by $\Theta_\infty$. Thus, we have
\[
  \Theta_\Gamma = \Theta_0 + \sum_{\mathbf r} \Theta_{\mathbf r} +
  \Theta_\infty\ ,
\]
where 
\begin{equation}
  \Theta_0 = -(fdU + gdV + h_ido^i)_{\rho=0}
\label{*}
\end{equation}
and 
\begin{equation}
  \Theta_\infty = \lim_{\rho=\infty}(fdU + gdV + h_ido^i) - {\mathcal
  N}_\infty E_\infty dt\ .
\label{**}
\end{equation}
The shortest way to calculate $\Theta_\Gamma$ found as yet uses the DNEF
gauge; this is, however, a singular gauge (cf.\ Sec.~3 of I), and some
justification is in order.

In Sec~3.2.2 of II, gauge transformation of the functions $F$, $G$ and
$H_i$ in the formulae (\ref{6,1})--(\ref{6,4}) have been calculated
starting from the requirement that the Liouville form remains
invariant. The invariance is in fact pointwise, that is, the integrand
of the Liouville form in any volume part is invariant at any point
$\rho$. Calculating the form in a singular gauge such as DNEF
coordinates then makes sense. At the points where the gauge is regular
the integrand has the same value as for a $C^1$ gauge. The DNEF gauge
is singular where the embedding intersects a horizon; the value of the
integrand at such a point can be defined as the limit from the left or
from the right because the integrand is continuous in a regular gauge.

This argument shows that the difference of the boundary terms obtained from
the integration over any volume part is gauge invariant and can be calculated
in the DNEF gauge. The only point at which caution is necessary is the
expression of the shell variables $\mathbf r$ and $\mathbf p$ in terms of
embedding variables and Dirac observables because the embeddings are not
continuous at the shell. Let us, therefore, generalize formulae (30) and (31)
of II,
\begin{eqnarray}
  {\mathbf p}_{\text{out}} & = & -R({\mathbf r})\Delta_{\mathbf
  r}(R_{,U})U'({\mathbf r})\ , 
\label{88.1}  \\ 
  {\mathbf p}_{\text{in}} & = & R({\mathbf r})\Delta_{\mathbf
  r}(R_{,V})V'({\mathbf r})\ . 
\label{88.2}
\end{eqnarray}
for the momentum and calculate the corresponding formulae for $\mathbf
r$. Again, the idea is that the value of $\mathbf p$ and $\mathbf r$ is gauge
invariant and we just have to express it in the singular gauge. 

Let us start with ${\mathbf p}_{\text{out}}$. Let us label the volume parts of
$\Sigma$ adjacent to the shell by the index $K = l,r$, $l$ meaning left and
$r$ meaning right from the shell. The corresponding pieces of Schwarzschild
spacetimes are denoted by ${\mathcal M}_K$, the Schwarzschild masses by $M_K$
and the maximal extensions of ${\mathcal M}_K$ by $\overline{\mathcal M}_K$.
We denote the singular gauge within $\overline{\mathcal M}_K$ by $U_K$ and
$V_K$. The coordinates $U$ and $V$ that occur in Eqs.~(\ref{88.1}) and
(\ref{88.2}) represent a $C^1$ gauge (see Sec.~2.2 of II). Hence, we have
\begin{multline*}
  {\mathbf p}_{\text{out}} = -R({\mathbf r})\left[\left(\frac{\partial
  R}{\partial U}\right)_r - \left(\frac{\partial R}{\partial
  U}\right)_l\right]U'({\mathbf r}) \\
  = -R({\mathbf r})\left[\left(\frac{\partial
  R}{\partial U}\right)_r\left(\frac{\partial
  U}{\partial \rho}\right)_r - \left(\frac{\partial R}{\partial
  U}\right)_l\left(\frac{\partial
  U}{\partial \rho}\right)_l \right] \\
  = -R({\mathbf r})\left[\left(\frac{\partial
  R}{\partial U_r}\frac{\partial
  U_r}{\partial \rho}\right)_{\rho={\mathbf r}} - \left(\frac{\partial
  R}{\partial U_l}\frac{\partial
  U_l}{\partial \rho}\right)_{\rho={\mathbf r}} \right],
\end{multline*}
and the generalized equation reads
\begin{equation}
  {\mathbf p}_{\text{out}} = -R({\mathbf r})\Delta_{\mathbf r}(R_{,U}U')\ .
\label{1c.1}
\end{equation}
Similarly,
\begin{equation}
  {\mathbf p}_{\text{in}} = R({\mathbf r})\Delta_{\mathbf r}(R_{,V}V')\ .
\label{1c.2}
\end{equation}

As for $\mathbf r$, we have to express it in terms of the coordinates of the
shell with respect to the singular gauge. Thus, we have immediately
\begin{eqnarray}
  U_K({\mathbf r}) & = & u_K\ ,
\label{1d.1} \\
  V_K({\mathbf r}) & = & v_K\ ,
\label{1d.2} 
\end{eqnarray}
where $u_K$ and $v_K$, $K = r,l$ are the new variables to describe the
position of the shell. As the functions $U_K(\rho)$ and $V_K(\rho)$ are
monotonous, the relation between $\mathbf r$ and any of $u_K$ and $v_K$ is
well defined for each $K = l,r$. Of course, the four parameters $u_l$, $v_l$,
$u_r$ and $v_r$ are redundant for the description of the shell position that
is determined by just one parameter $\mathbf r$, but the Liouville form which 
results at the end of the calculation contains only one combination of the four
parameters.

If the shell is marginally bound, that is, if it lies at a Schwarzschild
horizon of both spacetimes left and right, then either $u_K$ or $v_K$
diverges. However, since the Liouville form is gauge invariant, we obtain the
``proper'' value for it at these points of phase space as follows. First, we
calculate the Liouville form in a singular gauge for all cases but for the
marginally bound ones. Second, we transform it into some regular coordinates
on the phase space. Third, we take the limits to the marginally bound
cases. This will be done in Sec.~\ref{sec:C}.

\subsection{Contribution from a single shell}
\label{sec:single}
Let us calculate $\Theta_{\mathbf r}$ in the case that $p$, $\rho = {\mathbf
  r}$, is an intersection point of $\Sigma$ with a single shell. We assume
that the shell does not lie at the Schwarzschild horizons of
$\overline{\mathcal M}_K$ for $K = l,r$. Let the embedding defining $\Sigma$
be described by the pair of functions $(U_K(\rho), V_K(\rho))$ in ${\mathcal
  M}_K$ for each $K$. Then all contributions to $\Theta_{\mathbf r}$ are (cf.\ 
Eqs.~(51)--(55) of II):
\begin{eqnarray}
{\Theta}_{\mathbf r} &=& - {\Delta}_{\bf r}\Big[ f dU + g dV + h_i do^i +
  {\varphi} d{\bf  r } \Big] + {\bf p} d{\bf r} = 
\nonumber
\\
&& - {\Delta}_{\bf r}\bigg[ \bigg( {1 \over 2} \ln\left(-\frac{U'}{V'}\right)
  R R_{,U} + F \bigg)  \big( dU + U' d{\bf r}  \big)  \bigg] 
\nonumber
\\
&& - {\Delta}_{\bf r}\bigg[ \bigg( {1 \over 2} \ln\left(-\frac{U'}{V'}\right)
  R R_{,V} + G \bigg) \big( dV + V' d{\bf r}   \big)  \bigg] 
\nonumber
\\
&& - {\Delta}_{\bf r}\bigg[ \bigg( {1 \over 2} \ln\left(-\frac{U'}{V'}\right)
  R R_{,i} + H_i \bigg) do^i  \bigg] \nn \\  
&& + {\Delta}_{\bf r}\Big[  R R_{,U} U' d{\bf r} - R R_{,V} V' d{\bf r}  \Big] 
+ {\bf p} d{\bf r}  \, . 
\label{eq:boundary theta}
\end{eqnarray}
The symbol $\Delta_{\mathbf r}$ (that is defined in I, below Eq.~(19))
contains the indices $l$ and $r$ implicitly.

Now, we can start simplifying and transforming the right-hand side of
Eq.~(\ref{eq:boundary theta}). The first step is based on the properties of
the function $R$ along the shell: it is continuous, and it is defined, from
both sides, by
\begin{equation}
R({\bf r}) := R_K\big( U_K({\bf r}) , V_K({\bf r}) , o^i \big)\ .   
\label{eq:dR}
\end{equation}
Differentiating this equation, we obtain the relation
\begin{multline}
d\big(R({\bf r})\big) = R_{K,U_K}({\bf r}) \Big( dU_K({\bf r}) + U_K'({\bf r})
d{\bf r}  \Big) \\ + R_{K,V_K}({\bf r})  \Big( dV_K({\bf r}) + V_K'({\bf r})
d{\bf r} \Big) + R_{K,i}({\bf r}) do^i\ . 
 \label{eq:dR2}
\end{multline}
Moreover, taking differentials of Eqs.~(\ref{1d.1}) and (\ref{1d.2}) yields
\begin{equation}
d\big(U_K({\bf r})\big) = dU_K({\bf r}) + U'_K({\bf r}) d{\bf r} = du_K 
\label{eq:dr}
\end{equation}
and
\begin{equation}
d\big(V_K({\bf r})\big) = dV_K({\bf r}) + V'_K({\bf r}) d{\bf r} = dv_K \ .
\label{eq:dr2}
\end{equation}
Hence, Eq.~(\ref{eq:boundary theta}) simplifies to
\begin{eqnarray}
{\Theta}_{\mathbf r} &=& - {\Delta}_{\bf r}\bigg[ {1 \over 4}
\ln\left(-\frac{U'}{V'}\right)  d\big(R^2\big) + F du + G dv +H_i do^i
\bigg] 
\nonumber
\\
&& + {\Delta}_{\bf r}\bigg[  R R_{,U} U' d{\bf r} - R R_{,V} V' d{\bf r}
\bigg]  
+ {\bf p} d{\bf r}  \, . 
\label{eq:boundary THETA}
\end{eqnarray}

Further transformations depend on whether the shell at $p$ is out- or
in-going. Let us assume that it is out-going; the procedure for the in-going
shells is analogous.

In the next step, we calculate the value of the jumps that occur in
Eq.~(\ref{eq:boundary THETA}). The idea is that the embedding $\Sigma$ is
$C^1$; if the transformation between our singular gauge and a $C^1$ gauge were
known, one could compute the jumps. However, such a transformation is outlined
by Lemma 2 of I. At a point of an out-going shell that does not lie at a
Schwarzschild horizon (either a crossing of a shell and a horizon or a shell
lying at a horizon), we can apply the Lemma with the result: The
transformation of a regular gauge $U$ and $V$ to $U_l$ and $V_l$ is
\begin{equation}
  U = U_l\ ,\quad V = V_l
\label{2a.7}
\end{equation}
left from the shell, and to $U_r$ and $V_r$, it is determined by
\begin{equation}
  U = U_r - u_r + u_l
\label{2a.9}
\end{equation}
and 
\begin{equation}
  R_r(u_r,V_r(V),M_r) = R_l(u_l,V,M_l)\ ,
\label{2a.10}
\end{equation}
where $R_K(U_K,V_K,M_K)$ expresses the radius coordinate $R$ in the spacetime
${\mathcal M}_K$ as a function of the DNEF coordinates $U_K$ and $V_K$ and the
mass $M_K$, $K = l,r$.

Let us further recall that we admit only $C^1$ embeddings or else the
variational principle in II does not lead to the desired equations of motion
(cf.\ Sec.~2 of II). Hence, the derivatives $U'$ and $V'$ of the embedding
functions $U(\rho)$ and $V(\rho)$ must be continuous across the shell.  By
differentiating Eq. (\ref{2a.10}) with respect to $V$ right from the
shell we find that
\begin{equation}
 \frac{dV_r(V)}{dV} = \frac{R_{l,V_l}\big( u_l,V, M_l
   \big)}{R_{r,V_r}\big( u_r,V_r(V),M_r \big) }  \ .
\label{jump}
\end{equation}
Eq.~(\ref{jump}) implies that for all points $V(\rho)$ right from the shell we
have
\begin{equation}
 R_{r,V_r}(\rho) \, {V'}_r(\rho) = R_{l,V}\big( u_l,V(\rho), M_l
 \big) \, V'(\rho) \ .
\label{jump2}
\end{equation}
Eq. (\ref{2a.7}) implies for all points $V(\rho)$ left from the shell:
\begin{equation}
 R_{l,V}(\rho) \, {V'}_l(\rho) = R_{l,V}(\rho) \, V'(\rho) \ .
\label{jump3}
\end{equation}
We learn from
Eqs. (\ref{jump2}) and (\ref{jump3}) that  
\begin{equation}
 R_{r,V_r}({\bf r}) \, {V'}_r({\bf r}) 
= R_{l,V_l}({\bf r}) \, {V'}_l({\bf r}) 
\label{jump4}
\end{equation}
and, therefore, since $R(\rho)$ is continuous at $\rho={\bf r}$, that
\begin{equation}
{\Delta}_{\bf r}\Big[  R R_{,V} V' \Big] = 0\ . 
\label{jump5}
\end{equation}
Moreover, Eq.~(\ref{1c.1}) gives 
\begin{equation}
{\Delta}_{\bf r}\Big[  R R_{,U} U' \Big] + {\mathbf p} = 0\ .
\label{jump5'}
\end{equation}

Next, let us consider the discontinuity in the logarithm in
Eq. (\ref{eq:boundary THETA}) which  can be expressed as 
\begin{equation}
{\Delta}_{\bf r}\bigg[ \ln\left(-\frac{U'}{V'}\right) \bigg] = \ln\left(\frac{
    {U'}_r {V'}_l}{ {V'}_r {U'}_l}\right) \ . 
\label{jump6}
\end{equation}
Since $U(\rho)$ is continuous at $\rho={\bf r}$, we find from
Eqs. (\ref{2a.7}) and (\ref{2a.9}) that 
\begin{equation}
{U'}_r({\bf r}) = {U'}_l({\bf r}) \ .
\label{jump7}
\end{equation}
Eqs. (\ref{jump4}) and (\ref{jump7}) can be used to simplify (\ref{jump6}) and
express it in terms of the derivatives $R_{,V}$ at $\rho={\bf r}$,   
\begin{equation}
{\Delta}_{\bf r}\bigg[ \ln\left(-\frac{U'}{V'}\right) \bigg] = \ln\left(\frac{
    R_{r,V_r}({\bf r}) }{ R_{l,V_l}({\bf r}) }\right) \ . 
\label{jump8}
\end{equation}

The derivative $R_{K,V_K}$ as well as its sign can be determined in terms of
the radius coordinate $R_K$ and the mass $M_K$ if the position of the point
$p$ is further specified. It can lie in different quadrants of the extended
Schwarzschild spacetimes $\overline{\mathcal M}_K$. The quadrants have been
labeled by sign pairs $(\alpha,\beta)$ in Sec.~2.1 of I. The values of
$\alpha$ and $\beta$ also determine which DNEF coordinates are well defined in
the quadrant $(\alpha,\beta)$: they are $U^\alpha$ and $V^\beta$.

Suppose that the intersection $p$ lies in the quadrant $(\alpha_K,\beta_K)$ of
the Schwarzschild spacetime $\overline{\mathcal M}_K$. We must have $\alpha_r
= \alpha_l$ because of the argument of matching divergences of Sec.~2.2 in
I. Then the dependence of the function $R_K$ on the coordinates $U_K$ and
$V_K$ is given by Eq.~(5) of I:
\begin{equation}
  R = 2M_K\kappa\left[
  \alpha_K\beta_K\exp\left(
  \frac{-\alpha_KU_K^{\alpha_K} + \beta_KV_K^{\beta_K}}{4M_K}
  \right)\right]\ , 
\label{RK}
\end{equation}
where $\kappa$ is the Kruskal function (cf.\ Eq.~(6) of I). From the
definition of $\kappa$, the following equations result
\[
  (\kappa(x) - 1)e^{\kappa(x)} = x
\]
and
\[
 \kappa'(x) = \frac{1}{\kappa(x)e^{\kappa(x)}}\ .
\]
Using these equations to calculate $R_{K,V_K}$, we obtain
\begin{equation}
  \frac{\partial R_K}{\partial V_K^{\beta_K}} = \frac{{\beta_K}}{2} \bigg(1 -
  \frac{2M_K}{R_K} 
  \bigg) \ . 
\label{RV}
\end{equation}
We know that the expression
\[
  \alpha\beta\left(1 - \frac{2M}{R}\right)
\]
is always positive. Since $\alpha$ is the same from both sides, the
expression on the right-hand side of Eq.~(\ref{RV}) has the same sign
for both values of $K$ and
\[
  \ln\left(\frac{\partial R}{\partial V_r^{\beta_r}}\frac{\partial
  V_l^{\beta_l}}{\partial R}\right) = \ln\frac{|1 - 2M_r/R|}{|1 - 2M_l/R|} 
  = \Delta_{\mathbf r}\left(\ln\left|1 - \frac{2M}{R}\right|\right) =
  \Delta_{\mathbf r}\left(\ln|R - 2M|\right)\ .
\]
Thus, using Eqs.~(\ref{jump5}), (\ref{jump5'}) and (\ref{jump8}), we obtain
\begin{equation}
{\Theta}_{\mathbf r} = -{1 \over 4} d\big(R^2({\bf r})\big){\Delta}_{\bf
  r}(\ln| R -2M | )  
- {\Delta}_{\bf r}( F du + G dv + H_i do^i ) \ .
\label{eq:boundary THETA2'}
\end{equation}
Formula (\ref{eq:boundary THETA2'}) holds at each intersection of an out-going
shell with $\Sigma$ that is not a point of Schwarzschild horizon, $R =
2M_K$. Indeed, the first term diverges at such an intersection.

In the next step, we use the solutions for the functions $F$, $G$ and $H_i$
that have been found in Sec.~3.3 of II: in the DNEF gauge,
\begin{eqnarray}
F_K &=& 0 \ ,
\label{augas} \\ 
G_K &=& 0 \ ,
\label{augas2} \\ 
H_{Ki} &=&  -  \big( {\alpha_K} U_K^{\alpha_K} + {\beta_K} V_K^{\beta_K}
\big) \ 
\frac{{M_{K,i}}}{2}   \ ,
\label{augas3}
\end{eqnarray}
$K = l,r$. Using the formulae (\ref{1d.1}) and (\ref{1d.2}), we find that
\begin{equation}
{\Theta}_{\mathbf r} = -{1 \over 4} d\big(R^2({\bf r})\big){\Delta}_{\bf
  r}(\ln| R -2M | )  
+ \frac{1}{2} {\Delta}_{\bf r}\bigg[ \big( {\alpha} u + {\beta}
v \big) dM  \bigg] \ . 
\label{eq:boundary THETA2''}
\end{equation}

Eq.~(\ref{RK}) and the definition of $\kappa$ (Eq.~(6) of I)
imply
\begin{equation}
  \frac{-\alpha_Ku_K + \beta_Kv_K}{2} = R + 2M_K\ln\left|\frac{R}{2M_K} -
    1\right|\ .
\label{R*}
\end{equation}
Substituting this for $\beta_Kv_K$, we can cast Eq.~(\ref{eq:boundary
  THETA2''}) as follows:
\begin{equation}
{\Theta}_{\mathbf r} = {\Delta}_{\bf r}\bigg[ {\alpha} u^{\alpha}  dM - {1
  \over 4} 
d\big(R^2\big)  \ln\big| R -2M \big| +  R dM +
2M\ln\left|\frac{R - 2M}{2M}\right| dM \bigg] \ . 
\label{eq:boundary THETA2'''}
\end{equation}
For the last two terms in the bracket, however, the following identity can be
easily shown to hold:
\begin{multline}
  -{1\over 4} d\big(R^2\big)  \ln\big| R -2M \big| +  R dM +
  2M\ln\left|\frac{R - 2M}{2M}\right|dM \\ = d\left[-{1\over 4}(R^2 -
  4M^2)\ln|R  - 2M| - M^2\ln(2M) + {1\over 8}R^2 + {1\over 2}MR\right]\ .
\label{ident}
\end{multline}
Subtracting the corresponding exact form from $\Theta_{\mathbf r}$ leaves us
with 
\begin{equation}
{\Theta}_{\mathbf r} = {\Delta}_{\bf r}\Big[ {\alpha} u  dM \Big] =
{\alpha}_r \ u_r  \ dM_r -  {\alpha}_l  \ u_l\ dM_{m} \ . 
\label{eq:finalTHETA1}
\end{equation}

For an ingoing shell we find in an analogous way that
\begin{equation}
{\Theta}_{\mathbf r} = {\Delta}_{\bf r}\Big[ {\beta} v  dM \Big] = {\beta}_r
\ v_r  \ dM_r -  {\beta}_l  \ v_l  \ dM_l \ . 
\label{eq:finalTHETA2}
\end{equation} 
These two formulae summarize our treatment of single shells. Observe that the
singularity in $\Theta_{\mathbf r}$ due to shell crossing a horizon at $p$ has
disappeared: Eq.~(\ref{eq:finalTHETA2}) is well defined everywhere at any
shell that does not lie at a horizon.

\subsection{Contribution from a crossing point of two shells}
Let us next consider the case in which the embedding passes through the
crossing point $p$ of an in-going with an out-going shell. The form
$\Theta_{\mathbf r}$ at the crossing point $\rho = {\bf r}$ is given by
modified Eq.~(\ref{eq:boundary theta}), where we have to write ${\mathbf
  p}_{\text{out}}d{\mathbf r} + {\mathbf p}_{\text{in}}d{\mathbf r}$ instead
of ${\mathbf p}d{\mathbf r}$, and where the operator ${\Delta}_{\bf r}$ has a
different meaning: it does not denote jumps across one but across two shells.
This ought to be explained in more details.  The intersection of the two
shells defines four spacetime regions around the crossing point. Let us denote
them by ${\cal M}_K$, $K=l,r,u,d$, where the subscripts stand for left, right,
up, down. The embedding passes from the spacetime region on the left, ${\cal
  M}_l$, to the one on the right, ${\cal M}_r$, without entering in the upper
or lower intermediate regions. The operator ${\Delta}_{\bf r}$ in (\ref{eq:TWO
  THETA}) thus refers to the jump from ${\cal M}_l$ to ${\cal M}_r$. Let us
choose the coordinates in each region ${\cal M}_K$ to be the corresponding
DNEF coordinates $U_K$ and $V_K$.  The shells are described by
Eqs.~(\ref{1d.1}) and (\ref{1d.2}) but now $K = r,l,u,d$. Hence, the
coordinates $u_K, v_K$ in ${\cal M}_K$ satisfy
Eqs.~(\ref{eq:dr})--(\ref{eq:dr2}) for all $K$.

Since $R$ as a function on spacetime is continuous even at shell crossings,
Eq.~(\ref{eq:dR}) still holds, now for all four values of $K=l,r,d,u$. Hence,
Eq.~(\ref{eq:dR2}) also holds, in particular for $K=l$ and $r=r$, so we can
again collect all derivatives of $R$ as it has been done in
Sec.~\ref{sec:single}. The result is analogous to Eq.~(\ref{eq:boundary
  THETA}): 
\begin{eqnarray}
{\Theta}_{\mathbf r} &=& - {\Delta}_{\bf r}\bigg[ {1 \over 4}
\ln\left(-\frac{U'}{V'}\right)  d\big(R^2\big) + F du + G dv +H_i do^i
\bigg] 
\nonumber
\\
&& + {\Delta}_{\bf r}\bigg[  R R_{,U} U' d{\bf r} - R R_{,V} V' d{\bf r}
\bigg]  
+ {\bf p}_{\text{out}} d{\bf r} +  {\bf p}_{\text{in}} d{\bf r} \, . 
\label{9a'1}
\end{eqnarray}
We can even use Eqs.~(\ref{1c.1}) and (\ref{1c.2}) for the momenta with
${\Delta}_{\bf r}$ meaning the jump across the two shells. Indeed, let us
start with a regular gauge so that Lemma 1 of II is applicable. On the Lemma,
the jump in the $U$-derivative across an out-going shell is continuous along
the shell and it is zero across the in-going one. As the jump of any quantity
across two shells is the sum of jumps across each, we have the required
property, and we can proceed with the transformation to the singular gauge as
in the case on a single shell. Hence, we obtain the formula:
\begin{equation}
{\Theta}_{\mathbf r} \ = \ - {\Delta}_{\bf r}\bigg[ {1 \over 4}
  \ln\left(-\frac{U'({\bf r})}{V'({\bf r})}\right)  d\big(R^2({\bf r})\big) +
  F({\bf r}) du + G({\bf r}) dv +H_i({\bf r}) do^i    \bigg] \ . 
\label{eq:TWO THETA}
\end{equation}

Let us foliate a neighborhood of the crossing point by a $C^1$ family of
embeddings.  The foliation is described in ${\cal M}_K$ by $U_K(\tau,\rho)$,
$V_K(\tau,\rho)$ where the crossing point corresponds to $\tau={\bf t}$,
$\rho={\bf r}$; i.e., $U_K({\bf t},{\bf r})=u_K$ and $V_K({\bf t},{\bf
  r})=v_K$. Let the foliation intersect the ingoing shell at $\rho={\bf
  r}_{\rm in}(\tau)$ and the outgoing shell at $\rho={\bf r}_{\rm out}(\tau)$
so that ${\bf r}_{\rm in}({\bf t})={\bf r}={\bf r}_{\rm out}({\bf t})$. In
particular, this means that if $\tau \leq {\bf t}$ then $U_K(\tau,{\bf r}_{\rm
  out}(\tau))=u_K$ for $K=l,d$ and $V_K(\tau,{\bf r}_{\rm in}(\tau))=v_K$ for
$K=d,r$, while if $\tau \geq {\bf t}$ then $V_K(\tau,{\bf r}_{\rm
  in}(\tau))=v_K$ for $K=l,u$ and $U_K(\tau,{\bf r}_{\rm out}(\tau))=u_K$ for
$K=u,r$.

To calculate the jump in the logarithm term on the right-hand side of
Eq.~(\ref{eq:TWO THETA}), we again apply Lemma 2 of I. At each point of the
four ``legs''' of the crossing (trajectories of the shells outside $p$), we
can find $C^1$ coordinates and obtain equations for the jumps in $V'$ and $U'$
analogous to Eqs.~(\ref{jump4}) and (\ref{jump7}).  Therefore, taking the
limit ${\tau}={\bf t}$, we obtain equations like (\ref{jump4}) and
(\ref{jump7}) for all four spacetime regions around the crossing point:
\begin{eqnarray}
R_{r,V}({\bf r}) \, {V'}_{r}({\bf r}) = R_{u,V}({\bf r}) \, {V'}_u({\bf r}) \
, \ \  \ \ \ {U'}_{r}({\bf r}) = {U'}_u({\bf r}) \, , 
\label{j1}
\\
R_{l,V}({\bf r}) \, {V'}_{l}({\bf r}) = R_{d,V}({\bf r}) \, {V'}_d({\bf r}) \
, \ \  \ \ \ {U'}_{l}({\bf r}) = {U'}_d({\bf r}) \, , 
\label{j2}
\\
R_{u,U}({\bf r}) \, {U'}_{u}({\bf r}) = R_{l,U}({\bf r}) \, {U'}_l({\bf r}) \
, \ \ \ \ \ {V'}_{u}({\bf r}) = {V'}_l({\bf r}) \, , 
\label{j3}
\\
R_{r,U}({\bf r}) \, {U'}_{r}({\bf r}) = R_{d,U}({\bf r}) \, {U'}_d({\bf r}) \
, \ \ \ \ \ {V'}_{r}({\bf r}) = {V'}_d({\bf r}) \, . 
\label{j4}
\end{eqnarray}

Computing $R_{K,V}$ and $R_{K,U}$ from Eq.~(\ref{RK}) as for the single shell,
we can cast Eqs. (\ref{j1})--(\ref{j4}) in the following convenient form:
\begin{eqnarray}
\frac{{V'}_{r}({\bf r})}{{V'}_u({\bf r})} = \frac{ {\beta}_u  \big( R({\bf r})
  -2 M_u    \big)  }{ {\beta}_r \big( R({\bf r}) -2 M_r   \big)}\ ,  
\ \ \ \ \ \frac{{U'}_{r}({\bf r})}{{U'}_u({\bf r})} = 1 \ ,
\label{j11}
\\
\frac{{V'}_{l}({\bf r})}{{V'}_d({\bf r})} = \frac{ {\beta}_d  \big( R({\bf r})
  -2 M_d    \big)  }{ {\beta}_l \big( R({\bf r}) -2 M_l   \big)}\ ,  
\ \ \ \ \ \frac{{U'}_{l}({\bf r})}{{U'}_d({\bf r})} = 1 \ ,
\label{j22}
\\
\frac{{U'}_{u}({\bf r})}{{U'}_l({\bf r})} = \frac{ {\alpha}_l  \big( R({\bf
    r}) -2 M_l    \big)  }{ {\alpha}_u \big( R({\bf r}) -2 M_u   \big)}\ ,  
\ \ \ \ \ \frac{{V'}_{u}({\bf r})}{{V'}_l({\bf r})} = 1 \ , 
\label{j33}
\\
\frac{{U'}_{d}({\bf r})}{{U'}_r({\bf r})} = \frac{ {\alpha}_r  \big( R({\bf
    r}) -2 M_r    \big)  }{ {\alpha}_d \big( R({\bf r}) -2 M_d   \big)}\ ,  
\ \ \ \ \ \frac{{V'}_{d}({\bf r})}{{V'}_r({\bf r})} = 1 \ .
\label{j44}
\end{eqnarray}

Let us view them as algebraic equations determining the derivatives $U'$ and
$V'$ in terms of $R({\bf r})$, the masses $M_l$, $M_r$, $M_u$, $M_d$ and the
various indices $\alpha$ and $\beta$. Then it is easy to see that not all of
them are independent. Multiplying the equations for $U'$ in Eqs.~(\ref{j33})
and (\ref{j44}) and using the $U'$-equations from (\ref{j11}) and (\ref{j22})
lead to the condition
\begin{equation}
\frac{ {\alpha}_l  \big( R({\bf r}) -2 M_l    \big)  }{ {\alpha}_u \big(
  R({\bf r}) -2 M_u   \big)} = \frac{ {\alpha}_d \big( R({\bf r}) -2 M_d
  \big)}{ {\alpha}_r  \big( R({\bf r}) -2 M_r    \big)  } \, , 
\label{j111}
\end{equation}
while an analogous procedure for $V'$-equations yields 
\begin{equation}
\frac{ {\beta}_r \big( R({\bf r}) -2 M_r   \big)}{ {\beta}_u  \big( R({\bf r})
  -2 M_u    \big)  } = \frac{ {\beta}_d  \big( R({\bf r}) -2 M_d    \big)  }{
  {\beta}_l \big( R({\bf r}) -2 M_l   \big)} \, . 
\label{j222}
\end{equation}
Let us now recall the argument of matching divergence of Sec.~2.2 in I which
implies that not all of the indices $\alpha$ and $\beta$ are independent. In
general, two regions on opposite
sides of an outgoing shell should have the same $\alpha$, and two regions on
opposite sides of an ingoing shell should have the same $\beta$. In our case,
this means that ${\alpha}_u = {\alpha}_r$, ${\alpha}_l = {\alpha}_d$,
${\beta}_u = {\beta}_l$ and ${\beta}_r = {\beta}_d$. Therefore, Eqs.
(\ref{j111}) and (\ref{j222}) both reduce to the Dray-'t~Hooft-Redmount
condition \cite{DtH} and \cite{Red} (cf.\ Eq.~(7) of I),
\begin{equation}
\frac{  R({\bf r}) -2 M_l    }{ R({\bf r}) -2 M_u  } = \frac{ R({\bf r}) -2
  M_d  }{ R({\bf r}) -2 M_r    } \, , 
\label{DTR}
\end{equation}
determining the Schwarzschild radius $R({\bf r})$ of the crossing point in
terms of the masses of the four regions.  These variables are therefore not
all independent.

Let us now return to the logarithm term in Eq.~(\ref{eq:TWO THETA}) and use
Eqs.  (\ref{j11})--(\ref{j44}) to calculate its discontinuity from ${\cal
  M}_l$ to ${\cal M}_r$. The discontinuity
\begin{equation}
{\Delta}_{\bf r}\bigg[ \ln\left(-\frac{U'}{V'}\right) \bigg] = \ln\left(\frac{
    {U'}_{r}({\bf r}) {V'}_l({\bf r})}{ {V'}_{r}({\bf r}) {U'}_l({\bf
      r})}\right)  
\label{ole}
\end{equation}
can be expressed in various ways by using different pairs of equations from
the set (\ref{j11})--(\ref{j44}). Of course, because of the
Dray-'t~Hooft-Redmount condition (\ref{DTR}) all these alternative expressions
are equivalent.

The simplest and most symmetric of these is
\begin{equation}
 {\Delta}_{\bf r}\bigg[ \ln\left(-\frac{U'}{V'}\right) \bigg] = \ln\left|
  \frac{ R({\bf r}) -2 M_d }{R({\bf r}) -2 M_u} \right| \, ,
\label{ole2}
\end{equation}
which brings Eq.~(\ref{eq:TWO THETA}) to the form
\begin{eqnarray}
{\Theta}_{\mathbf r} \!\!\!&=&\!\!\! - {1 \over 4} d\big(R^2({\bf r})\big)
\ln | R({\bf r}) -2 M_d |  
+ {1 \over 4} d\big(R^2({\bf r})\big) \ln | R({\bf r}) -2 M_u |  
\nonumber
\\
&+&\!\!\! \frac{1}{2} \big( {\alpha}_r u_r + {\beta}_r v_r
\big) \ 
dM_r 
-  \frac{1}{2} \big( {\alpha}_l u_l + {\beta}_l v_l \big) 
 \ dM_l \ .
\label{eq:THETAA}
\end{eqnarray}
Notice that we have again chosen the particular solutions
(\ref{augas})--(\ref{augas3}) for the functions $F$, $G$ and $H_{i}$.  

The last two terms in Eq.~(\ref{eq:THETAA}) can be written with the help of
Eq. (\ref{R*}). Then, applying the identity (\ref{ident}), we obtain our final
expression for the contribution to the Liouville form from a crossing point,
involving the discontinuities ${\Delta}_{\bf r}$ from the left to the right
region and ${\Delta}_{\bf t}$ from the lower to the upper region:
\begin{eqnarray}
{\Theta}_{\mathbf r} & = & \frac{1}{2} {\Delta}_{\bf t}\bigg[  \big(-
{\alpha} u + 
{\beta} v \big) \ dM   \bigg] + \frac{1}{2} {\Delta}_{\bf r}\bigg[
\big( {\alpha} u + {\beta} v \big) \ dM  \bigg] \nn  \\
&&=  \frac{1}{2} \big(- {\alpha}_u u_u + {\beta}_u v_u \big)\ dM_u -   
\frac{1}{2} \big(- {\alpha}_d u_d + {\beta}_d v_d \big) \ dM_d  \nn  \\
&&+ \frac{1}{2} \big( {\alpha}_r u_r + {\beta}_r v_r \big) \ dM_r  
-  \frac{1}{2} \big( {\alpha}_l u_l + {\beta}_l v_l \big) 
 \ dM_l \ .
\label{eq:crossfinal}
\end{eqnarray}
Again, this formula holds only if no shell at the crossing is marginally
bound.

\subsection{The general formula for all contributions}
Before we conclude this section let us verify that the above formula for a
crossing point agrees with the previous formulae (\ref{eq:finalTHETA1}) and
(\ref{eq:finalTHETA2}) for ingoing and outgoing shells. Indeed, one can
consider an embedding starting from the left and then passing either above or
below or through a crossing point. If it passes above, it crosses first
the ingoing shell and then the outgoing shell. The total contribution from
both shells is the sum of the contributions (\ref{eq:finalTHETA1}) and
(\ref{eq:finalTHETA2}):
\begin{equation}
{\Theta}_{\text{above}} = {\beta}_{u}  \ v_{u}  \ dM_{u} -  {\beta}_l  \
v_{l}  \ dM_{l} + {\alpha}_{r} \ u_{r}  \ dM_{r} -
{\alpha}_u  \ u_{u}  \ dM_{u} \ . 
\label{upemb}
\end{equation}
If it passes below, it crosses the outgoing shell first and the ingoing
one second. By summing the contributions (\ref{eq:finalTHETA1}) and
(\ref{eq:finalTHETA2}) we now get 
\begin{equation}
{\Theta}_{\text{below}} = {\alpha}_{d} \ u_{d}  \ dM_{d} -  {\alpha}_l  \
u_{l}  \ dM_{l} + {\beta}_{r}  \ v_{r}  \ dM_{r} -  {\beta}_d
\ v_{d}  \ dM_{d}  \ . 
\label{loemb}
\end{equation}
Expressions (\ref{upemb}) and (\ref{loemb}) are equivalent, and their
symmetric sum indeed recovers the result (\ref{eq:crossfinal}) obtained by
letting the embedding pass through the crossing point. We can prove the
equivalence as follows.

The difference ${\Theta}_{\text{above}} - {\Theta}_{\text{below}}$ can be
written as 
\[
  {\Theta}_{\text{above}} - {\Theta}_{\text{below}} = -\chi_l dM_l + \chi_d
  dM_d + \chi_u dM_u - \chi_r dM_r\ ,
\]
where
\[
  \chi_K := -\alpha_Ku_K + \beta_Kv_K\ .
\]
Eq.~(\ref{R*}) allows us to express $\chi$'s in terms of the four masses
$M_K$,
\begin{equation}
  \chi_K = 2R({\mathbf r}) + 4M_K\ln|R({\mathbf r}) - 2M_K| - 4M_K\ln(2M_K)\ ,
\label{12a.2}
\end{equation}
if we use formula (\ref{DTR}) to calculate $R({\mathbf r})$:
\begin{equation}
  R({\mathbf r}) = 2\frac{M_lM_r - M_dM_u}{M_l + M_r - M_d - M_u}\ .
\label{12a.1}
\end{equation}
In this way ${\Theta}_{\text{above}} - {\Theta}_{\text{below}}$ is a well
defined form on the four-dimensional space spanned by $M_l$, $M_r$, $M_d$ and
$M_u$. Observe that, from the point of view of canonical transformations, our
``old momenta'' are $M_l$, $M_r$ and $M_d$, while the ``new momenta''
are $M_l$, $M_r$ and $M_u$ and the form ${\Theta}_{\text{above}} -
{\Theta}_{\text{below}}$ is the differential of a generating function $\mathcal
F$ that depends on the old and new coordinates, if we can show that it is
an exact form. Let us show that.

First, we observe that ${\Theta}_{\text{above}} - {\Theta}_{\text{below}}$ is
invariant with respect to the swaps 
\begin{equation}
  M_r \longleftrightarrow M_l\ ,\quad M_d \longleftrightarrow M_u\ .
\label{swaps}
\end{equation}
Hence, it is sufficient to show that
\[
  -\frac{\partial \chi_l}{\partial M_d}  = \frac{\partial \chi_d}{\partial
  M_l}\ , \qquad  \frac{\partial \chi_l}{\partial M_r} = \frac{\partial
  \chi_r}{\partial  M_l}\ ,\qquad 
  \frac{\partial \chi_d}{\partial M_u} = \frac{\partial \chi_u}{\partial
  M_d}\ . 
\]

Second, we observe that $\chi_K$ depends on the masses with index different
from $K$ only through $R({\mathbf r})$. Hence the above equations can be
written as 
\begin{eqnarray}
  -\frac{\partial \chi_l}{\partial R({\mathbf r})}\ \frac{\partial R({\mathbf
  r})}{\partial M_d} & = & \frac{\partial \chi_d}{\partial R({\mathbf r})}\
  \frac{\partial R({\mathbf r})}{\partial M_l}\ , 
\label{12c.1} \\
  \frac{\partial \chi_l}{\partial R({\mathbf r})}\ \frac{\partial R({\mathbf
  r})}{\partial M_r} & = & \frac{\partial \chi_r}{\partial R({\mathbf r})}\
  \frac{\partial R({\mathbf r})}{\partial M_l}\ , 
\label{12c.2} \\
  \frac{\partial \chi_d}{\partial R({\mathbf r})}\ \frac{\partial R({\mathbf
  r})}{\partial M_u} & = & \frac{\partial \chi_u}{\partial R({\mathbf r})}\
  \frac{\partial R({\mathbf r})}{\partial M_d}\ .
\label{12c.3} 
\end{eqnarray}
Eq.~(\ref{12a.2}) yields
\begin{equation}
  \frac{\partial \chi_K}{\partial R({\mathbf r})} = \frac{2R({\mathbf
  r})}{R({\mathbf r}) - 2M_K}\ ,
\label{12c.4}
\end{equation}
and Eq.~(\ref{12a.1}) implies
\[
  \frac{\partial R({\mathbf r})}{\partial M_l} =
  2\frac{(M_r-M_d)(M_r-M_u)}{(M_l+M_r-M_d-M_u)^2}
\]
and 
\[
  \frac{\partial R({\mathbf r})}{\partial M_d} =
  2\frac{(M_u-M_l)(M_u-M_r)}{(M_l+M_r-M_d-M_u)^2}\ ;
\]
the other formulae can be obtained through the swaps (\ref{swaps}). We also
find that
\[
  R({\mathbf r}) - 2M_l = -2\frac{(M_l-M_d)(M_l-M_u)}{M_l+M_r-M_d-M_u}
\]
and
\[
  R({\mathbf r}) - 2M_d = 2\frac{(M_d-M_l)(M_d-M_r)}{M_l+M_r-M_d-M_u}\ ,
\]
the other two differences resulting by the swaps. It follows that
\begin{equation}
  \frac{\partial R({\mathbf r})}{\partial M_l} = -\frac{R({\mathbf r}) -
  2M_r}{M_l+M_r-M_d-M_u} 
\label{12d.1}
\end{equation}
and
\begin{equation}
  \frac{\partial R({\mathbf r})}{\partial M_d} = \frac{R({\mathbf r}) -
  2M_u}{M_l+M_r-M_d-M_u}\ .
\label{12d.2}
\end{equation}
Substituting Eqs.~(\ref{12c.4}), (\ref{12d.1}) and (\ref{12d.2}) or their
suitable swaps into Eqs.~(\ref{12c.1})--(\ref{12c.3}), we easily show the
equality if we use Eq.~(\ref{DTR}) once more.

Since the form ${\Theta}_{\text{above}} - {\Theta}_{\text{below}}$ is exact,
its integral along a curve depends only on the end points of the curve. Let us
calculate such an integral in the $M$-space. Let the curve start at the origin
and consist of four straight segments: $(0,0,0,0)\rightarrow(M_d,0,0,0)$,
$(M_d,0,0,0)\rightarrow(M_d,M_u,0,0)$,
$(M_d,M_u,0,0)\rightarrow(M_d,M_u,M_r,0)$ and
$(M_d,M_u,M_r,0)\rightarrow(M_d,M_u,M_r,M_l)$. An elementary integration
yields
\begin{multline}
  {\mathcal F} = -2M_lM_r + 2M_dM_u \\
  -2M^2_l\ln\frac{|M_l-M_d||M_l-M_u|}{M_l(M_l+M_r-M_d-M_u)} 
  -2M^2_r\ln\frac{|M_r-M_d||M_r-M_u|}{M_r(M_l+M_r-M_d-M_u)} \\
  +2M^2_d\ln\frac{|M_d-M_l||M_d-M_r|}{M_d(M_l+M_r-M_d-M_u)}
  +2M_u^2\ln\frac{|M_u-M_l||M_u-M_r|}{M_u(M_l+M_r-M_d-M_u)}\ . 
\label{10/1}
\end{multline}
Strictly speaking, this formula holds only for that part of the
$M$-space where the masses $M_d$ and $M_u$ are larger that $M_l$ and $M_r$
because the integration curve would otherwise cross the corresponding 
singularities of
the integrand. But one guesses that this formula holds in all generic subcases
as it stands. We can check this as follows.

The function $\mathcal F$ is to be a generating function of the canonical
transformation that brings the Liouville form $\Theta_{\text{below}}$ into
$\Theta_{\text{above}}$. It depends on the old $(M_l,M_d,M_r)$ and the new
$(M_l,M_u,M_r)$ momenta and the usual generating formulae must hold (as some
new momenta coincide with some old ones, differences of coordinates appear
instead of the coordinates themselves):
\begin{eqnarray}
  \frac{\partial {\mathcal F}}{\partial M_d} & = & -(\beta_dv_d-\alpha_du_d)\
  ,   
\label{11/1} \\
  \frac{\partial {\mathcal F}}{\partial M_u} & = & -(\beta_uv_u-\alpha_uu_u)\ ,
\label{11/2} \\
  \frac{\partial {\mathcal F}}{\partial M_l} & = & (\beta_lbv_l-\alpha_lu_l)\ .
\label{11/3} \\
  \frac{\partial {\mathcal F}}{\partial M_r} & = & (\beta_rbv_r-\alpha_ru_r)\ .
\label{11/3'}
\end{eqnarray}
A simple calculation shows that Eqs.~(\ref{11/1})--(\ref{11/3}) are valid in
all subcases with $a \neq 0$ and $b \neq 0$ for our guessed $\mathcal F$ as
given by Eq.~(\ref{10/1}).

The transformation itself can be calculated as follows. Let us introduce the
following notation
\begin{alignat*}{3}
  q_1 & = -\alpha_lu_l\ , & \qquad q_2 & = \beta_rv_r\ , & \qquad q_3 & =
  \alpha_du_d - \beta_dv_d\ , \\
  p_1 & = -M_l\ , & \qquad p_2 & = -M_r\ , & \qquad p_3 & = -M_d\ , \\  
  Q_1 & = -\beta_lv_l\ , & \qquad Q_2 & = \alpha_ru_r\ , & \qquad Q_3 & =
  -\alpha_uu_u + \beta_uv_u\ , \\
  P_1 & = -M_l\ , & \qquad P_2 & = -M_r\ , & \qquad P_3 & = -M_u\ .
\end{alignat*}
Then $\Theta_{\text{below}} = \sum-q_ndp_n$ and $\Theta_{\text{above}} =
\sum-Q_ndP_n$. Eq.~(\ref{R*}) for $K = d$ gives
\begin{equation}
  R({\mathbf r}) =
  -2p_3\kappa\left[\alpha_d\beta_d\exp\left(\frac{q_3}{4p_3}\right)\right] 
\label{3'.1}
\end{equation}
and Eq.~(\ref{12a.1}) yields
\begin{equation}
  P_3 = \frac{p_1p_2 - (p_1+p_2-p_3)p_3
  \kappa\left[\alpha_d\beta_d\exp\left(\frac{q_3}{4p_3}\right)\right]}{p_3 -  
  p_3\kappa\left[\alpha_d\beta_d\exp\left(\frac{q_3}{4p_3}\right)\right]}\ .
\label{3'.2}
\end{equation}
Using Eq.~(\ref{R*}) again, we obtain
\begin{equation}
  Q_3 = 2R({\mathbf r}) - 4P_3\ln\left|\frac{R({\mathbf r}) +
  2P_3}{-2P_3}\right|\ ,
\label{4'.1}
\end{equation}
where $R({\mathbf r})$ and $P_3$ are given by Eqs.~(\ref{3'.1}) and
(\ref{3'.2}). Then,
\begin{equation}
  P_1 = p_1\ ,\qquad P_2 = p_2
\label{4'.2}
\end{equation}
and Eq.~(\ref{R*}) implies 
\begin{eqnarray}
  Q_1 & = & q_1 - 2R({\mathbf r}) + 4p_1\ln\left|\frac{R({\mathbf r}) +
  2p_1}{-2p_1}\right|\ ,
\label{4'.3} \\
  Q_2 & = & q_2 - 2R({\mathbf r}) + 4p_2\ln\left|\frac{R({\mathbf r}) +
  2p_2}{-2p_2}\right|\ ,
\label{4'.4}
\end{eqnarray}
where again $R({\mathbf r})$ is to be substituted from Eq.~(\ref{3'.1}). The
desired transformation is described by Eqs.~(\ref{3'.1})--(\ref{4'.4}); it is
a rather involved one.
 
Finally, let us make the following observation. Let us view the operator
${\Delta}_{\bf r}$ as the jump in the positive spacelike direction and the
operator ${\Delta}_{\bf t}$ as the jump in the positive timelike direction.
This orientation of spacetime is fixed by our particular definition of DNEF
coordinates. Then, we see that along an outgoing shell ${\Delta}_{\bf r} = -
{\Delta}_{\bf t}$, and along an ingoing shell ${\Delta}_{\bf r} =
{\Delta}_{\bf t}$. By substituting these relations into expression
(\ref{eq:crossfinal}) we recover our previous expressions
(\ref{eq:finalTHETA1}) and (\ref{eq:finalTHETA2}) as special cases. Let us
therefore summarize all our results of this section by rewriting expressions
(\ref{eq:finalTHETA1}), (\ref{eq:finalTHETA2}) and (\ref{eq:crossfinal})
concisely as
\begin{equation}
{\Theta} = {\Delta}_{\bf t}\Big[  {R^*} \ dM   \Big] +  {\Delta}_{\bf r}\Big[
T \ dM  \Big] \ , 
\label{eq:ffinal}
\end{equation}
where
\[
  R^*_K := \frac{-\alpha_Ku_K+\beta_Kv_K}{2}\ ,
\]
and 
\[
  T_K := \frac{\alpha_Ku_K+\beta_Kv_K}{2}\ .
\]
This holds in general, with $\rho = {\bf r}$ corresponding to the point of
intersection of the shell(s) with the chosen embedding if none of the shells
is marginally bounded.

\subsection{Contributions from center and from infinity}
For the center and infinity, we shall again utilize the freedom in the gauge
choice and compute in the DNEF coordinates in the spacetime part surrounding
the origin, which we denote by${\cal M}_0$ and the part near infinity, which
will be denoted by ${\cal M}_\infty$. For simplicity we shall drop the
subscripts $0$ and $\infty$ from $U$ and $V$ in this section.

\subsubsection{At the center} 
The contributions to the Liouville from $\rho = 0$ are given by Eq.~(\ref{*}),
where the functions $f$, $g$ and $h_i$ are defined by Eqs.
(\ref{6,1})--(\ref{6,3}). The solution (\ref{augas}), (\ref{augas2}) and
(\ref{augas3}) for the functions $F$, $G$, $H_i$ is trivial in the Minkowski
part of spacetime,
\begin{equation}
F = 0 \, \; \; \; G = 0 \, \; \; \; H_i = 0 \, ,
\label{eq:trivial}
\end{equation}
and therefore Eq. (\ref{*}) reduces to an expression involving only
the logarithm terms in $f$, $g$ and $h_i$.  

At the regular center $R=0$ (where $\rho = 0$) all embeddings have to satisfy
the boundary condition  
\begin{equation}
U'(0) = - V'(0) \, ,
\label{eq:primed}
\end{equation}
which guarantees that the embedded hypersurfaces avoid conical singularities.
Eq.  (\ref{eq:primed}) reduces all logarithm terms in Eqs.
(\ref{6,1})--(\ref{6,3}) to zero. The functions $f$, $g$ and $h_i$ and
consequently the Liouville form at $\rho = 0$ is therefore trivial:
\begin{equation}
{\Theta}_0 = 0 \, .
\label{eq:center 0}
\end{equation}

\subsubsection{At infinity} 
The contributions to the Liouville from $\rho \rightarrow \infty$ are given by
Eq.~(\ref{**}). At infinity, we restrict the foliation to be parallel to the
$T = $ const, where $T$ is the Schwarzschild time coordinate in ${\mathcal
  M}_\infty$; more precisely, we assume that
\begin{equation}
R(\rho) \rightarrow \rho + O({\rho}^{-1}) \, , \; \; \;  T(\rho) \rightarrow
T_{\infty} + O({\rho}^{-1}) \ ,
\label{eq:TRinf}
\end{equation}
for the Schwarzschild coordinates $T$ and $R$ along any embedding. We also
assume that the foliation parameter $t$ satisfies
\[
  t \rightarrow T_\infty
\]
asymptotically so that ${\mathcal N}_\infty = 1$. Finally, we replace
$E_\infty$ by $M_\infty$, $M_\infty$ being the total mass of the system. Then
we can write
\begin{equation}
  {\Theta}_\infty = \lim_{\rho\rightarrow\infty}(f dU + g dV + h_i do^i)  -
  M_\infty dT_{\infty}\, . 
\label{eq:infinity}
\end{equation}

The functions ${f}$, ${g}$ and ${h}_i$ have always the same functional form,
given by Eqs.~(\ref{6,1})--(\ref{6,3}) and ${F}$, ${G}$ and ${H}_i$ are given
by Eqs.~(\ref{augas})--(\ref{augas3}), where $M_K$ is replaced by $M_\infty$.

Let us start from the only non-trivial contribution, Eq.~(\ref{augas3}),
and use the fact that ${V} + {U} = 2T$. As $\rho$ approaches
infinity the time coordinate $T$ approaches the asymptotic time $T_{\infty}$,
which means that the total contribution from ${H}_i$ is just
\begin{equation}
{H}_i do^i = -T_{\infty} dM_\infty \, .
\label{eq:Hcontrinf}
\end{equation}
This term together with the term $-M_\infty dT_{\infty}$ in Eq.
(\ref{eq:infinity}) yields an exact form. The remaining contributions to the
Liouville form in Eq.  (\ref{eq:infinity}) are proportional to the logarithm
terms and can be summarized as follows:
\begin{equation}
{\Theta}_\infty = \frac{{R}}{2}\ln\left(-\frac{{U}'}{{V}'}\right)
\Big( {R}_{,{U}} d{U} + {R}_{,{V}} d{V}  +
{R}_{,i} do^i  \Big) \, . 
\label{eq:infinity3}
\end{equation}

Recall that in a Schwarzschild region of mass $M$ the Eddington-Finkelstein
coordinates are related to the coordinates $T$ and $R$ by  
\begin{eqnarray}
{U} = T - R^* \, , \; \; \; {V} = T + R^* \, ,
\label{eq:U and V}
\\
R^* = R + 2M {\rm ln}\left|{\frac{R}{2M}}-1\right| \, . 
\label{eq:Rstar}
\end{eqnarray}
As $\rho \rightarrow \infty$, it follows from Eq. (\ref{eq:Rstar}) that $R^*$
has the asymptotic expansion 
\begin{equation}
R^*(\rho) \rightarrow \rho + 2M {\rm ln}\left( {\frac{\rho}{2M}} \right) +
O({\rho}^{-1}) \, . 
\label{eq:Rstarinf}
\end{equation}
Eq. (\ref{eq:U and V}) determines how the Eddington-Finkelstein coordinates
approach infinity, 
\begin{eqnarray}
{U}(\rho) \!\!&\rightarrow&\!\!  - \rho - 2M {\rm ln}\left(
  {\frac{\rho}{2M}} \right) + T_{\infty} + O({\rho}^{-1}) \, , 
\label{eq:Uinf}
\\
{V}(\rho) \!\!&\rightarrow&\!\!  + \rho + 2M {\rm ln}\left(
  {\frac{\rho}{2M}} \right) + T_{\infty} + O({\rho}^{-1}) \, , 
\label{eq:Vinf}
\end{eqnarray}
and hence how their derivatives ${U}'$ and ${V}'$ approach it:
\begin{eqnarray}
{U}'(\rho) \!\!&\rightarrow&\!\! - 1 + O({\rho}^{-1}) \, , 
\label{eq:U'inf}
\\
{V}'(\rho) \!\!&\rightarrow&\!\! 1 + O({\rho}^{-1})   \, .
\label{eq:V'inf}
\end{eqnarray}
The way in which the differentials $d{U}$ and $d{V}$ increase as
$\rho \rightarrow \infty$ follows directly from Eqs. (\ref{eq:Uinf}) and
(\ref{eq:Vinf}), 
\begin{eqnarray}
d{U}(\rho) \!\!&\rightarrow&\!\!  -2dM {\rm ln}\left(
  {\frac{\rho}{2M}} \right) + dT_{\infty}+ 2dM + O({\rho}^{-1}) \, , 
\label{eq:dU inf}
\\
d{V}(\rho) \!\!&\rightarrow&\!\!  +2dM {\rm ln}\left(
  {\frac{\rho}{2M}} \right) + dT_{\infty}- 2dM + O({\rho}^{-1}) \, , 
\label{eq:dV inf}
\end{eqnarray}
while the way in which the logarithm in Eq. (\ref{eq:infinity3}) behaves is
determined from Eqs. (\ref{eq:U'inf}) and (\ref{eq:V'inf}): 
\begin{equation}
\ln\left(-\frac{{U}'}{{V}'}\right)(\rho) \rightarrow
O({\rho}^{-1})   \, . 
\label{eq:ln inf}
\end{equation}

The final terms whose behavior we need to determine are the terms multiplying
the logarithm in Eq. (\ref{eq:infinity3}). Eqs. (\ref{RK}) and the fact that
we are in the $(++)$ quadrant imply that the derivatives of ${R}$ are related
to ${R}$ and $M$ by
\begin{eqnarray}
{{R}}_{,{U}} = - {1 \over 2} \left( 1 - {\frac{2M}{{R}}}
\right) = - {{R}}_{,{V}} \; , 
\nonumber
\\
{{R}}_{,i} = {\frac{{{R}}}{M}}M_{,i} - {1 \over 2} \left( 1 -
  {\frac{2M}{{R}}} \right) {\frac{{V} -  
  {U} }{M}} M_{,i} \; .  
\label{eq:Rderiv}
\end{eqnarray}
After some simple calculations it follows that
\begin{eqnarray}
{R}_{,{U}}(\rho) \rightarrow -{1 \over 2} + {M \over \rho} +
O({\rho}^{-3}) \, , \; \; \; \; {R}_{,{V}}(\rho) \rightarrow {1
  \over 2} - {M \over \rho} + O({\rho}^{-3}) \, , 
\nonumber
\\
{R}_{,i}(\rho) \rightarrow 2M_{,i} - 2M_{,i} {\rm ln}\left(
  {\frac{\rho}{2M}} \right) + O\big({\rho}^{-1} {\rm ln}(\rho)\big) \, . 
\label{eq:Rderivinf}
\end{eqnarray}
Putting together Eqs. (\ref{eq:dU inf}), (\ref{eq:dV inf}) and
(\ref{eq:Rderivinf}) it is not difficult to show that  
\begin{equation}
\Big( {R}_{,{U}} d{U} + {R}_{,{V}} d{V}  +
{R}_{,i} do^i  \Big)(\rho) \rightarrow O\big({\rho}^{-1} {\rm
  ln}(\rho)\big) \,  .
\label{eq:allRderivinf}
\end{equation}
When Eqs. (\ref{eq:ln inf}) and (\ref{eq:allRderivinf}) are used in
Eq. (\ref{eq:infinity3}) they yield the final result; the total contribution
to the Liouville form from infinity vanishes: 
\begin{equation}
\Theta_\infty = 0\, .  
\label{eq:finalresult}
\end{equation}

\section{Algebra of Dirac's observables}
\label{sec:algebra}
All calculations done and all results obtained as yet in this paper have been
entirely general in the sense that they hold for a system containing any
number of out- and in-going null spherical shells. In the present section, we
return to our original system of two shells.  Our final aim is to express the
Liouville form in terms of a chosen set of Dirac observables for all cases
studied in I and so to find the symplectic structure of the physical phase
space. In the previous section, the Liouville form has been reduced to the sum
over contributions from shells that intersect a Cauchy surface. Each
intersection of the surface either with single shells or with crossing points
of shells contributes according to the general formula (\ref{eq:ffinal}).

\subsection{The parallel shells: Cases A and B}
The subspacetimes between the shells are denoted by ${\cal M}_K$, where the
index $K=l,m,r$ (see Fig.~1 in I). The spacetime region ${\cal M}_l$ is flat
and lies entirely in the $(+,+)$ quadrant. The masses of the other two regions
are $M_m$ and $M_r$.  The DNEF coordinates of each region ${\cal M}_K$ are
denoted by $U_K$, $V_K$. The regions ${\cal M}_m$ and ${\cal M}_r$ lie in the
$(+,+)\cup(-,+)$ part of their corresponding Schwarzschild extensions. The two
shells are characterized by the index $s=1,2$. The position of the $s$-shell
with respect to the chart $U_K$, $V_K$ covering the region ${\cal M}_K$ is
defined by $V^+_K=v^+_{Ks}$. By applying Eq.~(\ref{eq:ffinal}) to case A we
find the following total Liouville form:
\begin{equation}
  {\Theta}_\Gamma = v_{m1} \ dM_{m} - v_{m2} \ dM_{m}   + v_{r2}  \
  dM_{r} \ . 
\label{LfA}
\end{equation}
It is clear from (\ref{LfA}) that the momentum conjugate to the coordinate
$v_{r2}$ is $-M_r$ and the momentum conjugate to the difference $v_{m1}-
v_{m2}$ is $-M_m$. These four coordinates describe the physical phase space
of the system, which is the subset of ${\rm I}\!{\rm R}^4$ defined by the
inequalities
\begin{equation}
  M_m > 0 \ , \ \ \ M_r > M_m \ .
\label{PhspAi}
\end{equation}
The coordinate $v_{r2}$ can take any values between $-\infty$ and $+\infty$,
while $v_{m1} - v_{m2}$ is negative in case A and positive in case A'.

Different coordinates can be defined by the canonical transformation
\begin{alignat*}{2}
  M_{(1)} & = M_m\ , & \quad M_{(2)} & = M_r - M_m \ , \\
  v_{(1)} & = v_{m1} - v_{m2} + v_{r2} \ , 
  & \quad v_{(2)} & = v_{r2}\ .
\end{alignat*}
These coordinates are associated more closely to the individual shells:
$M_{(1)}$ is the mass of the first shell and $M_{(2)}$ is that of the
second. Moreover, the coordinate $v_{(1)}$ can be considered as the advanced
time of the first shell with respect to the continuous time coordinate at the
past null infinity. That is, we can choose a gauge so that $v_{m2} =
v_{r2}$ and the coordinate $V$ is continuous across the shell.

In terms of these coordinates the physical phase space is the subset of ${\rm
  I}\!{\rm R}^4$ defined by 
\begin{equation}
  M_{(1)} > 0 \ , \ \ \ M_{(2)} > 0 \ .
\label{PhspAii}
\end{equation}
The variables $v_{(2)}$ and $v_{(1)}$ are free to take any values between
$-\infty$ and $+\infty$ but $v_{(2)} > v_{(1)}$ in case A while $v_{(2)} <
v_{(1)}$ in case A'.  In case B and B', the subspacetimes are again denoted by
${\cal M}_K$, $K=l,m,r$, and the masses of the two non-flat regions are $M_m$
and $M_r$. The left region lies in the $(+,+)$ quadrant of the flat spacetime
while the other two in the $(+,+)\cup(+,-)$ part of their Schwarzschild
extension. The position of the $s$-shell is defined by $U^+_K=u_{Ks}$. Using
(\ref{eq:ffinal}) we find that the total Liouville form is
\begin{equation}
  {\Theta}_\Gamma = u_{m1} \ dM_{m} - u_{m2} \ dM_{m}   + u_{r2}  \
  dM_{r} \ . 
\label{LfB}
\end{equation}
The description of the phase space is similar to the description above, and
analogous choices of natural physical coordinates can be made.

\subsection{The crossing shells: Cases C or C', and subcases}
\label{sec:C}
Cases C and C' are obtained from each other by interchanging the values $1,2$
of the label $s$. So we can consider only one of these cases. Let us also drop
$s$ since the positions of the shells are already distinguished by the letters
$u$ and $v$. The four spacetime regions are denoted by ${\cal M}_K$ where
$K=l,r,u,d$. The various subcases are pictured in Figs.~4--9 in I. The
spacetime region ${\cal M}_l$ is flat and the masses of the other regions are
$M_r$, $M_u$ and $M_d$. The left region lies in the $(+,+)$ quadrant. The
other regions may lie in a union of two or four quadrants of their
Schwarzschild extension depending on whether each shell lies above, exactly
on, or below its corresponding horizon. The various possibilities are captured
by simple relationships between the masses $M_r$, $M_u$ and $M_d$.  For this
reason, the indices $a$ and $b$ have been introduced in I, defined by
\begin{equation}
a := {\rm sgn}(M_r - M_u) \ , \ \ \ \ b := {\rm sgn}(M_r - M_d) \ .
\label{ablabels}
\end{equation}
Their meaning is the following: Because of the flatness of the left region the
only shell boundaries that can lie on horizons are the boundaries between
${\cal M}_d$ and ${\cal M}_r$ and between ${\cal M}_u$ and ${\cal M}_r$. If a
shell lies above the horizon (corresponding to an unbound state) the
difference between the Schwarzschild mass on its right and that on its left is
positive, if it lies below the horizon (bound state) this difference is
negative, and if it lies exactly on the horizon (marginally bound state) it is
zero. The indices $a$ and $b$ hence provide all the necessary information and
characterize the subcases ${\rm C}_{ab}$ by the nine combinations of their
values $(+,0,-)$.

The values of the signs $\alpha_K$ and $\beta_K$ for each case C$_{ab}$ have
been found in I: the relations 
\[
  \alpha_l = +1\ ,\quad \beta_l = +1\ ,\quad \alpha_d = +1\ ,\quad \beta_u =
  +1 
\]
hold for all cases; if $a \neq 0$,
\[
  \alpha_u = \alpha_r = a
\]
and if $b \neq 0$,
\[
  \beta_d = \beta_r = b\ .
\]
If $a=0$ the coordinates $u_u$ and $u_r$ diverge, and if $b=0$ $v_d$ and
$v_r$ diverge.

Let us consider the cases with $a \neq 0$: C$_{++}$, C$_{+0}$, C$_{+-}$,
C$_{-+}$, C$_{-0}$ and C$_{--}$. The coordinates $v_l$, $v_u$, $u_u$ and
$u_r$ are regular, and we can write the form $\Theta_{\text{above}}$ as
follows: 
\begin{equation}
  \Theta_{\text{above}} = (-au_u + v_u)dM_u + au_rdM_r\ . 
\label{19a.1}
\end{equation}
This region of the phase space can, therefore, be considered as a Darboux chart
with the coordinate pairs 
\begin{equation}
  -au_u + v_u,-M_u\ ;\quad  au_r,-M_r\ .
\label{19b.1}
\end{equation}

Next, consider the cases with $b \neq 0$: C$_{++}$, C$_{0+}$, C$_{-+}$,
C$_{+-}$, C$_{0-}$ and C$_{--}$. Here, $u_l$, $u_d$, $v_d$ and
$v_r$ are regular, and $\Theta_{\text{below}}$ is
\begin{equation}
  \Theta_{\text{below}} = (u_d - bv_d)dM_d + bv_rdM_r\ .
\label{19b'.2}
\end{equation}
Hence, this region of the phase space is covered by the Darboux chart with
coordinate pairs
\begin{equation}
  -u_d - bv_d,-M_d\ ;\quad  bv_r,-M_r\ .
\label{19b.2}
\end{equation}

The two charts overlap in C$_{++} \cup \text{C}_{-+} \cup \text{C}_{+-} \cup
\text{C}_{--}$. The (canonical) transformation between the two charts in the
overlapping region is given by the formulae (\ref{3'.1}), (\ref{3'.2}),
(\ref{4'.1}), (\ref{4'.2}) and (\ref{4'.4}), where we have to set:
\begin{equation}
  q_2 = bv_r\ ,\quad q_3 = u_d - bv_d\ ,\quad p_2 = -M_r\ ,\quad p_3 = -M_d 
\label{19c.1}
\end{equation}
and 
\begin{equation}
  Q_2 = au_r\ ,\quad Q_3 = -au_u + v_u\ ,\quad P_2 = -M_r\ ,\quad P_3 = -M_u \
  , 
\label{19c.2}
\end{equation}
and $p_1 = P_1 = 0$. Let us write down the transformation explicitly:
\begin{eqnarray}
  Q_2 & = & q_2 + 4p_3\bar{\kappa} +
  4p_2\ln\left|\frac{p_2-p_3\bar{\kappa}}{p_2}\right|\ ,   
\label{21a.1} \\
  Q_3 & = & - 4p_3\bar{\kappa} - \frac{4\bar{\kappa}}{\bar{\kappa}-1}(p_2-p_3)
  \ln\left|\frac{p_2-p_3\bar{\kappa}}{p_2-p_3}\right|\ ,
\label{21a.2} \\
  P_2 & = & p_2\ ,
\label{21a.3} \\
  P_3 & = & \frac{\bar{\kappa}}{\bar{\kappa}-1}(p_2-p_3)\ ,
\label{21a.4} 
\end{eqnarray}
where $\bar{\kappa}$ is a shorthand for 
\begin{equation}
  \bar{\kappa} = \kappa\left[b\exp\left(\frac{q_3}{4p_3}\right)\right]
\label{21a.5}
\end{equation}
and the coordinates $q_1$ and $Q_1$ do not occur in the formulae. The
generating function ${\mathcal F}'$ for the transformation is given by
Eq.~(\ref{10/1}) if we set $M_l = 0$ in it:
\begin{multline*}
  {\mathcal F}' = 2M_uM_d - 2 M_r^2\ln\frac{|M_r-M_d||M_r-M_u|}{M_r(M_r - M_d
  - M_u)} \\ 
  + 2M_d^2\ln\frac{|M_r - M_d|}{M_r - M_d - M_u} + 2M_u^2\ln\frac{|M_r -
  M_u|}{M_r - M_d - M_u}\ .
\end{multline*}
The charts cover the whole phase space with the exception of case C$_{00}$. 

To find coordinates that are regular at C$_{00}$, we first use
Eq.~(\ref{4'.4}) rewritten in terms of the three masses,
\[
  Q_2 = q_2 - \frac{4M_dM_u}{M_d+M_u-M_r} -
  4M_r\ln\left|\frac{(M_r-M_d)(M_r-M_u)}{M_r(M_d+M_u-M_r)}\right|
\]
and consider the fact that $Q_2$ diverges at $M_u = M_r$ while $q_2$ does when
$M_d = M_r$. It follows that the combination
\[
  Q_2 + 4M_r\ln|M_r-M_u|
\]
is always regular. 

Let us introduce a new coordinate $q$ by
\begin{equation}
  Q_2 = q - \frac{4M_dM_u}{M_d+M_u-M_r} -
  4M_r\ln\left|\frac{M_r-M_u}{M_d+M_u-M_r}\right|\ .
\label{21b.1}
\end{equation}
This should be regular; the motivation for addition of further terms is that
\begin{equation}
  \bar{v}_u = q - 4(M_r-M_u)\ln\left|\frac{M_r-M_u}{M_d+M_u-M_r}\right|\ ,
\label{21b.2}
\end{equation}
hence, $q \rightarrow \bar{v}_u$ for $M_u \rightarrow M_r$. Here, the
coordinate $\bar{v}_u$ is defined by $\bar{v}_u := Q_2 + Q_3$; we would have
$\bar{v}_u = v_u$ if we shift $v_u$ and $u_u$ so that $u_u = u_r$. If we
express $Q_3$ in Eq.~(\ref{4'.1}) in terms of masses,
\begin{equation}
  Q_3 = \frac{4M_dM_u}{M_d+M_u-M_r} +
  4M_u\ln\left|\frac{M_r-M_u}{M_d+M_u-M_r}\right|\ ,
\label{21c.2}
\end{equation}
we can see that Eq.~(\ref{21b.2}) holds. The choice of $\bar{v}_u$ instead of
$q$ does not, however, lead to differentiable components of the Liouville
form.

Let us view Eqs.~(\ref{21b.1}) and (\ref{21c.2}), together with $P_2 = -M_r$
and $P_3 = -M_u$ as transformation to new variables $q$, $M_d$, $M_u$ and
$M_r$. Then
\begin{multline*}
  \Theta_\Gamma = - Q_2dP_2 - O_3dP_3 \\
  = qdM_r + \frac{2M_d(M_r-M_u)}{M_d+M_u-M_r}d(M_r-M_u) -
  \frac{2(M_r^2-M_u^2)}{M_d+M_u-M_r}dM_d \\
  + d\left[-2(M^2_r-M_u^2)\ln\left|\frac{M_r-M_u}{M_d+M_u-M_r}\right|\right]\
  .
\end{multline*}
If we subtract the singular exact form, the rest seems to be regular. But we
have also to calculate the determinant of the symplectic tensor in order to
show the regularity. Some simplification can be achieved in the coordinates
$q$, $x$, $y$ and $z$ defined by
\[
  x = M_r - M_u\ ,\quad y = M_r + M_u\ ,\quad z = M_d\ .
\]
The corresponding components of the symplectic form $\Omega = d\Theta_\Gamma$
are 
\begin{alignat*}{3}
  \Omega_{qx} & = 1/2\ , & \quad \Omega_{qy} & = 1/2\ , & \quad \Omega_{qz} &
  =  0\ , \\
  \Omega_{xy} & = 0\ , & \quad  \Omega_{xz} & = 2\frac{x^2-yz}{(z-x)^2}\ , &
  \quad \Omega_{yz} & = 2\frac{x^2-xz}{(z-x)^2} 
\end{alignat*}
and the determinant is
\[
  \det{\mathbf \Omega} = \frac{z^2}{(z-x)^4}(x-y)^2 =
  \frac{4M_d^2M_u^2}{(M_d+M_u-M_r)^4}\ .
\]

The C-component ${\mathcal P}_{\text{C}}$ of the physical phase space is
determined by the boundaries (cf.~I, Eqs.~(9) and (10))
\[
  M_d > 0\ ,\quad M_u > 0\ ,
\]
and
\[
  0 < M_r < M_d + M_u\ .
\]
Hence, $\Omega$ is $C^\infty$ and non-degenerate everywhere in ${\mathcal
  P}_{\text{C}}$. In particular, case C$_{00}$, which is defined by $M_d = M_u
= M_r$, is a smooth surface in ${\mathcal P}_{\text{C}}$.

\section{Conclusions and outlook}
\label{sec:outlook} 
In the three papers I, II and the present one, complete sets of Dirac
observables for the system of two null-matter shells have been found and their
Poisson bracket have been determined. The result is exceedingly simple and
analogous to what was found earlier for the single shell \cite{H-Kie}, but the
calculation itself, in spite of many improvements, has still been tedious.  We
don't give up hopes that methods exist to make the calculation truly simpler.

Let us turn to the question of quantization. It seems that the similarity of
the present results to those of \cite{H-Kie} suggests that at least some
quantization methods can be imported from \cite{H}. More specifically,
there does not seem to be any reason why each shell of the pair could not
bounce at the regular center exactly as the single shell did in
\cite{H}. Some singularities that afflict the classical solutions could so be
again avoided by the quantum theory.

On the other hand, the pair of shells is crucially different from the single
shell in one aspect: the shells can intersect and they have a non-trivial
interaction at the intersection. The transformation
(\ref{21a.1})--(\ref{21a.5}) describes the change of the (initial) Dirac
observables before into the (final) ones after the crossing. It can so
represent the dynamical evolution through the crossing and it ought to be
derivable from the Hamiltonian of the system. It may be difficult to find such
a Hamiltonian on account of the transformation being rather involved. Perhaps
there is another choice of Dirac observables that simplifies the
transformation. 

This is a technical problem. However, there is another very interesting aspect
of the crossing that may lead to a great change in results. This is the
fact that the external shell can become bound after the crossing. For the
values of the initial observables $q_3$ and $p_3$ such that
\[
  \kappa\left[\exp\left(\frac{q_3}{4p_3}\right)\right] \in
  \left(1,\frac{p_2}{p_3}\right)\ ,
\]
the Eqs.~(\ref{21a.3}) and (\ref{21a.4}) yield
\[
  M_r \leq M_u
\]
and a black hole forms, while for 
\[
  \kappa\left[\exp\left(\frac{q_3}{4p_3}\right)\right] \in
  \left(\frac{p_2}{p_3},\infty\right)
\]
the external shell can reach infinity. Here, we assume that $p_2/p_3 > 1$ so
$M_r > M_d$ and the external shell before the crossing is not bound. This
indicates that the pair of shells might only sometimes re-expand to infinity
and the other times form a ``quantum black hole''. Recall that the single
shell always re-expands and reach the infinity (cf.~\cite{H}). The possibility
that a black hole forms would, therefore, constitute a qualitatively new
result. We hope to be
able to construct the quantum theory soon.

\section*{Acknowledgments}
The authors are thankful for useful discussions by K.~V.~ Ku\-cha\v{r}. The
work has been supported by the Swiss Nationalfonds and by the Tomalla
Foundation, Zurich.

\end{document}